\documentclass[11pt]{article}

\usepackage[margin=1in]{geometry}
\usepackage[T1]{fontenc}
\usepackage[utf8]{inputenc}
\usepackage{lmodern}
\usepackage{hyperref}
\usepackage{booktabs}
\usepackage{graphicx}
\usepackage{array}
\usepackage{enumitem}
\usepackage{xcolor}
\usepackage{float}
\usepackage{tabularx}
\usepackage{subcaption}

\title{RepoReviewer: A Local-First Multi-Agent Architecture for Repository-Level Code Review}
\author{Peng Zhang \\ \texttt{pengz7720@gmail.com}}
\date{}

\begin{document}

\maketitle

\begin{abstract}
Repository-level code review requires reasoning over project structure, repository context, and file-level implementation details. Existing automated review workflows often collapse these tasks into a single pass, which can reduce relevance, increase duplication, and weaken prioritization. We present RepoReviewer, a local-first multi-agent system for automated GitHub repository review with a Python CLI, FastAPI API, LangGraph orchestration layer, and Next.js user interface. RepoReviewer decomposes review into repository acquisition, context synthesis, file-level analysis, finding prioritization, and summary generation. We describe the system design, implementation tradeoffs, developer-facing interfaces, and practical failure modes. Rather than claiming benchmark superiority, we frame RepoReviewer as a technical systems contribution: a pragmatic architecture for repository-level automated review, accompanied by reusable evaluation and reporting infrastructure for future empirical study.
\end{abstract}

\section{Introduction}

Large language models are increasingly used for code review, yet many practical review tasks extend beyond isolated code snippets. Repository-level review requires understanding project layout, configuration, documentation, and file interactions. In practice, single-pass prompting often mixes these concerns into one generation step, which can lead to noisy findings, weak severity ranking, and poor summaries.

This paper presents RepoReviewer, a practical multi-agent architecture for automated repository review. The system is designed as a local-first developer tool with three user-facing surfaces: a command-line interface, a web dashboard, and a backend API. The review workflow decomposes the task into specialized stages for repository acquisition, context building, file-level review, prioritization, and final reporting.

The contribution of this work is primarily systems-oriented. We do not introduce a new learning method or model architecture. Instead, we contribute a concrete multi-agent workflow for repository-level code review, a local-first implementation combining Python, FastAPI, LangGraph, PyGithub, LiteLLM, and Next.js, a documented implementation spanning CLI, API, and web UI surfaces, and a discussion of practical failure modes, tradeoffs, and directions for future evaluation.

\section{Problem Setting}

RepoReviewer takes as input either a public GitHub repository URL or a repository URL plus a pull request number. It produces structured review comments with file, line, severity, issue, suggestion, and snippet fields, a ranked summary of findings, and disk artifacts in JSON and Markdown formats.

The system is designed under several practical constraints. Repository contents may exceed model context limits, repositories contain generated, binary, or oversized files that should be excluded, useful review output must be prioritized and deduplicated, and the tool must operate across different provider APIs and models.

\section{System Overview}

\begin{figure}[t]
    \centering
    \includegraphics[width=\linewidth]{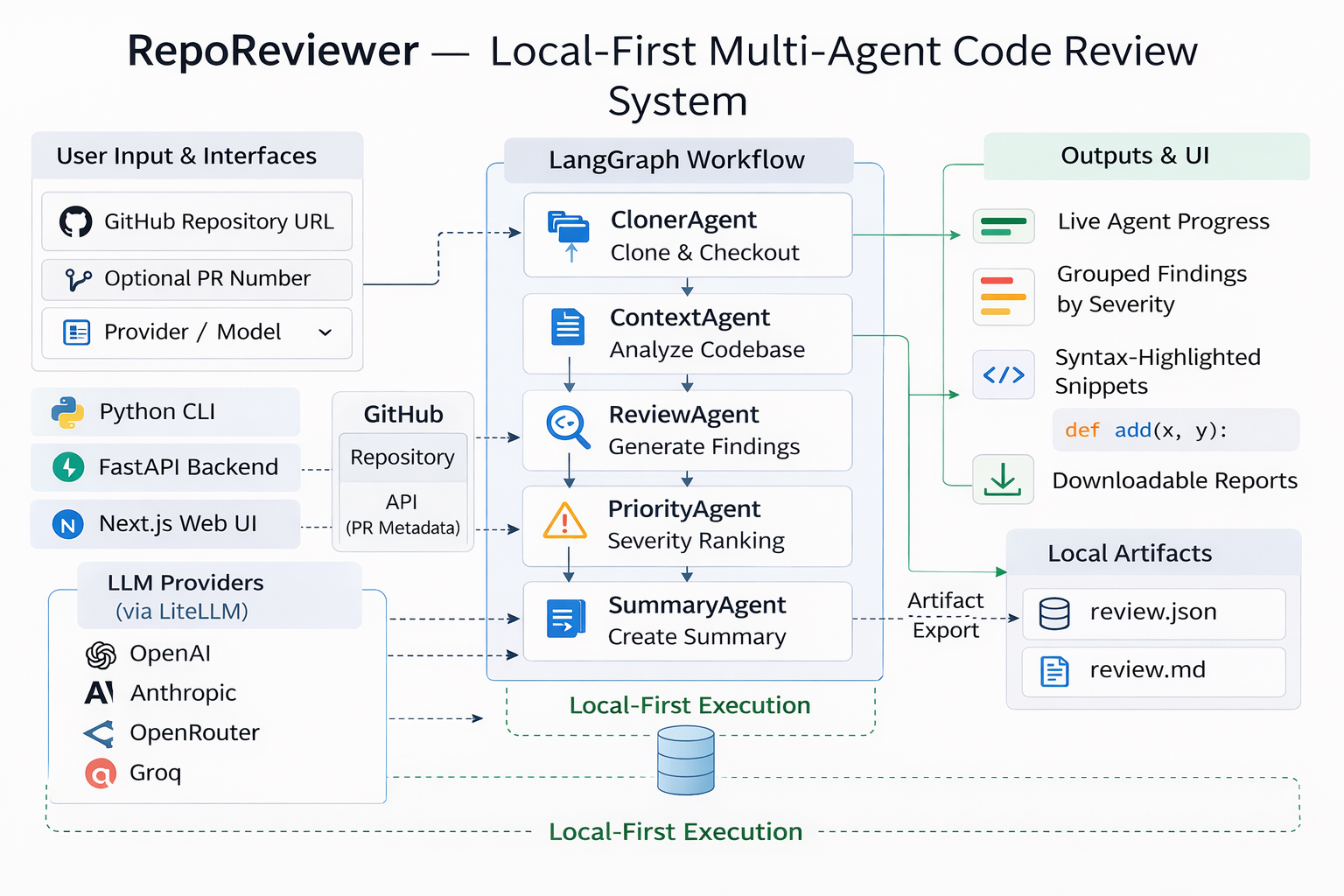}
    \caption{RepoReviewer system architecture showing user-facing interfaces, the LangGraph workflow, external provider and GitHub integrations, and local artifact generation.}
    \label{fig:architecture}
\end{figure}

Figure~\ref{fig:architecture} illustrates the RepoReviewer pipeline.

RepoReviewer uses a hybrid multi-agent design. Deterministic steps are used where explicit logic is more reliable than model generation, while language-model calls are reserved for context synthesis, code review, and final summarization.

\subsection{Design Principles}

The system is guided by four implementation principles. First, review jobs run on the developer's machine and write artifacts directly to disk, reflecting the project's local-first execution model. Second, repository acquisition, context synthesis, review, prioritization, and summarization are separated into distinct stages rather than folded into a monolithic prompt. Third, the backend exposes a provider-agnostic model interface through LiteLLM so that different hosted models can be used without changing the workflow design. Fourth, JSON and Markdown artifacts are treated as first-class outputs rather than optional logs.

\subsection{ClonerAgent}

The ClonerAgent clones a target repository locally and, for pull-request mode, resolves the PR head branch to ensure the reviewed content matches the requested change set.

\subsection{ContextAgent}

The ContextAgent extracts repository structure, README content, and key file previews, then synthesizes a project-level context summary. This stage is intended to reduce shallow file-by-file analysis by providing architectural cues to later steps.

\subsection{ReviewAgent}

The ReviewAgent iterates over selected files and generates structured review comments. For each comment, the system attempts to preserve file location, severity, issue description, suggestion, and a code snippet for rendering in the CLI and web UI.

\subsection{PriorityAgent}

The PriorityAgent deduplicates and sorts findings by severity. In the current implementation, prioritization is largely deterministic rather than fully model-driven.

\subsection{SummaryAgent}

The SummaryAgent generates a concise human-readable report that emphasizes top findings and skipped-file categories.

\section{Implementation}

The backend is implemented in Python and uses FastAPI for API delivery, LangGraph for workflow orchestration, LiteLLM for model-provider abstraction, and PyGithub for GitHub pull-request metadata. The frontend is implemented in Next.js and receives review progress through server-sent events.

The system exposes three usage modes: CLI review through a Typer-based command-line interface, API-triggered review jobs through FastAPI, and browser-based review initiation with progress visualization through the Next.js dashboard.

The tool is currently local-first. This decision reduces deployment overhead and simplifies access to generated artifacts, though it also constrains collaborative or hosted use.

\subsection{Component Boundaries}

The implementation is intentionally split across a shared review engine and thin interface layers. The Python core owns repository acquisition, file selection, provider calls, ranking, and artifact writing. The CLI invokes this core directly. The FastAPI service wraps the same core in background jobs and progress streams. The Next.js frontend acts primarily as a job launcher and result renderer rather than implementing review logic itself. This separation keeps review behavior consistent across interfaces and reduces the risk that one surface diverges from another.

\subsection{Output Artifacts}

Each review run produces \texttt{review.json} for structured downstream analysis, \texttt{review.md} for developer-readable reporting, grouped findings in the web UI with snippets and severity buckets, and intermediate experiment outputs when the evaluation runner is used. This artifact-first design makes the system easier to audit, easier to annotate for future studies, and easier to reuse outside the browser interface.

\section{Demonstration and Analysis}

Rather than presenting RepoReviewer as a benchmarked research system, we frame this work as a technical report on system design and implementation. The current evidence is therefore qualitative and engineering-focused. We use a combination of live system runs, generated artifacts, and interface demonstrations to analyze the behavior of the system.

\subsection{Demonstration Scenarios}

RepoReviewer supports three primary workflows: command-line review of a repository or pull request, API-triggered review jobs with artifact downloads, and browser-based review initiation with agent progress streaming and grouped result rendering.

The implementation has been exercised on public repositories using multiple provider configurations. Representative outputs include structured findings, ranked summaries, Markdown reports, JSON artifacts, and interface screenshots. The current report emphasizes these concrete artifacts because they reflect the system's operational behavior without overstating empirical certainty.

\begin{figure}[H]
    \centering
    \begin{subfigure}[t]{0.32\linewidth}
        \centering
        \includegraphics[width=\linewidth]{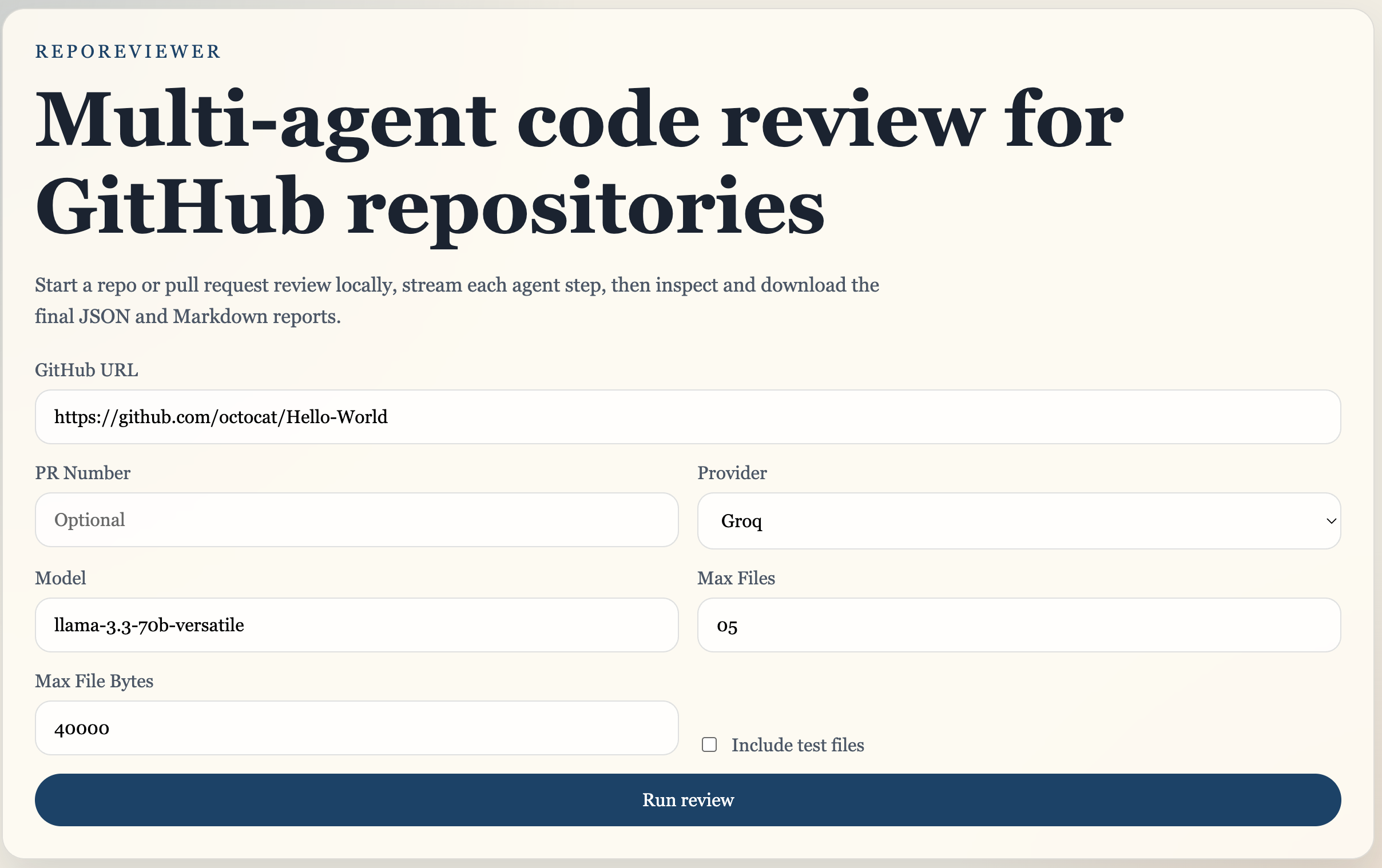}
        \caption{Home screen}
    \end{subfigure}
    \hfill
    \begin{subfigure}[t]{0.32\linewidth}
        \centering
        \includegraphics[width=\linewidth]{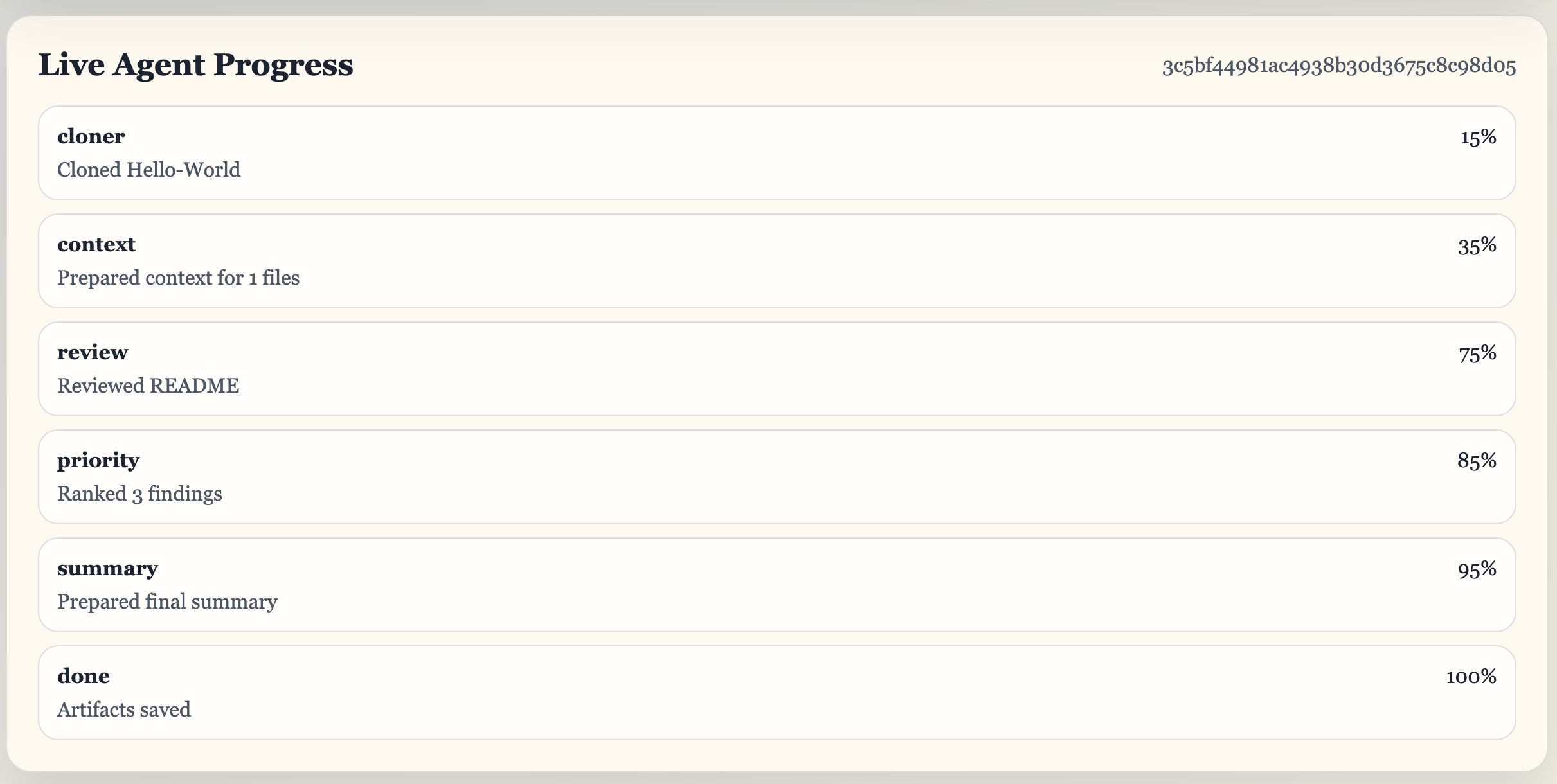}
        \caption{Live progress}
    \end{subfigure}
    \hfill
    \begin{subfigure}[t]{0.32\linewidth}
        \centering
        \includegraphics[width=\linewidth]{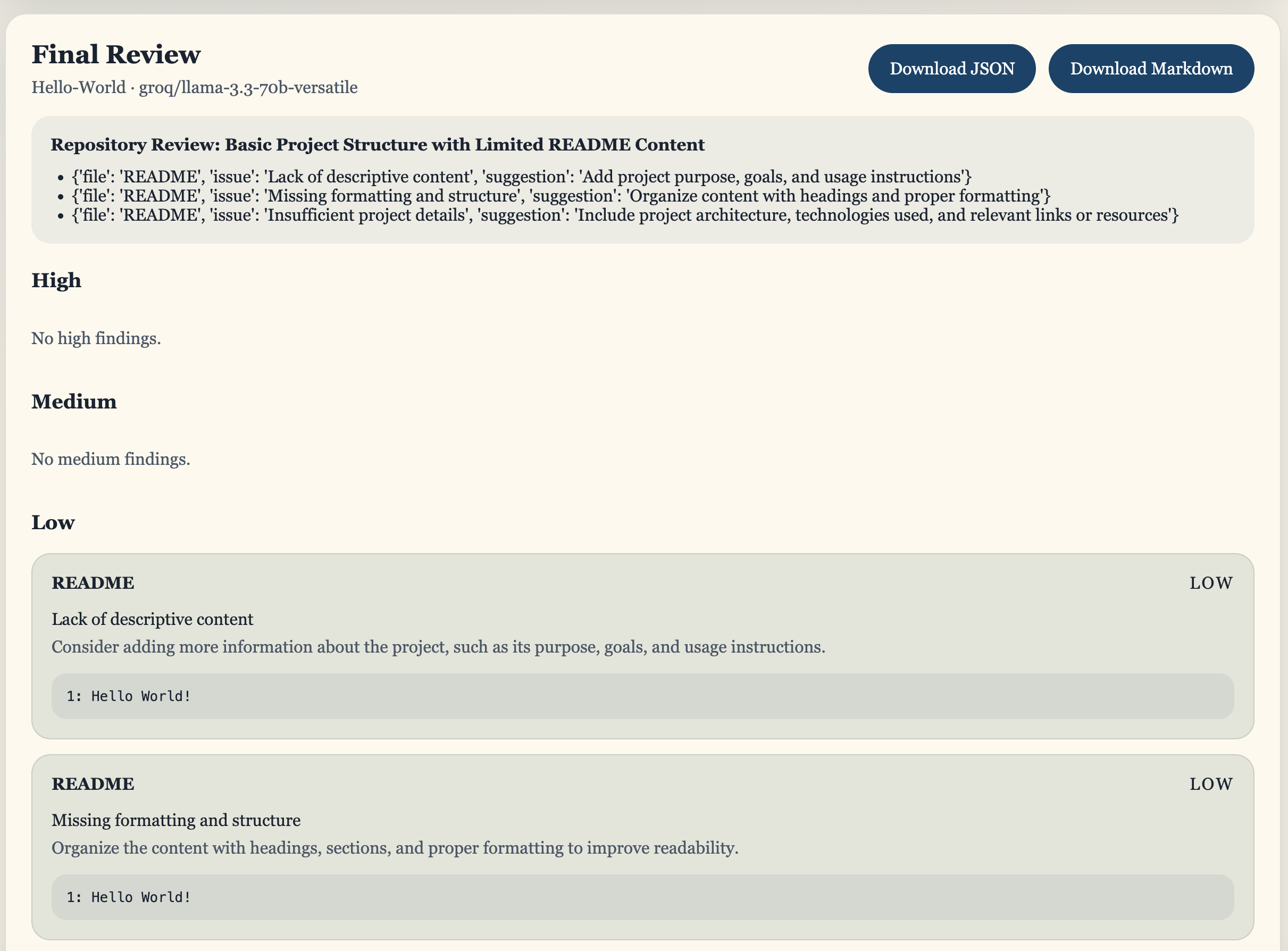}
        \caption{Final report}
    \end{subfigure}
    \caption{RepoReviewer web interface across job submission, staged agent execution, and final artifact rendering.}
    \label{fig:ui-panels}
\end{figure}

\subsection{Illustrative Examples}

The system can already generate complete outputs for small public repositories. For example, a minimal repository such as \texttt{octocat/Hello-World} produces repository-context summaries, file-level findings for the extensionless \texttt{README} file, and a final Markdown report. While this example is not sufficient for a benchmark claim, it demonstrates that the full pipeline is operational end to end.

The implementation also supports comparative runs across multiple review modes, including the full workflow, a single-agent baseline, and ablation variants. These modes are exposed through the evaluation runner and produce artifacts suitable for later annotation and aggregation. In this sense, the project already contains the mechanics needed for future empirical work even though the present paper stops at architecture and demonstration.

\subsection{What the Demonstration Shows}

The current implementation demonstrates that the end-to-end pipeline is operational across CLI, API, and UI surfaces, that repository context can be collected and passed into later reasoning stages, that structured findings can be ranked, rendered, and exported consistently, and that the architecture is extensible enough to support future ablations and evaluation studies. What the current report does \emph{not} demonstrate is statistically grounded superiority over simpler baselines. That claim is deliberately deferred until a larger human-annotated evaluation is completed.

\subsection{Observed Practical Constraints}

In practical usage, several system-level constraints emerge. Provider quota limits can interrupt larger batch evaluations, model output variability requires defensive parsing and type coercion, repository quality strongly affects review quality when documentation is sparse, and larger repositories force tradeoffs between context breadth, latency, and cost.

\section{Future Evaluation Plan}

Although this report does not claim completed benchmark results, the implementation includes an evaluation harness designed for future study. That harness supports full-system and single-agent comparison runs, ablations such as no-context and no-priority variants, export of per-run artifacts and summary tables, flattened annotation sheets for human review, and post-annotation aggregation into CSV, JSON, and LaTeX summaries. The intended evaluation metrics are precision, actionable rate, duplicate rate, severity agreement, top-ranked usefulness, runtime, and estimated cost. These metrics require human annotation and sufficient provider quota to execute a larger multi-repository study.

\section{Implementation-Supported Reporting Workflow}

The repository includes a paper workspace with a manuscript skeleton, an annotation rubric, dataset templates for review runs, a results CSV template, an evaluation runner, annotation aggregation utilities, and generated LaTeX table export for future result integration. This workflow is intended to reduce the engineering burden of future empirical study, even though the present report focuses on system documentation rather than completed quantitative comparison.

\begin{table}[h]
    \centering
    \begin{tabular}{lcccc}
        \toprule
        Asset & Purpose & Produced by & Current status \\
        \midrule
        JSON report & structured findings & review pipeline & implemented \\
        Markdown report & human-readable summary & review pipeline & implemented \\
        Annotation sheet & human evaluation input & evaluation runner & implemented \\
        LaTeX summary table & paper integration & annotation aggregator & implemented \\
        \bottomrule
    \end{tabular}
    \caption{Implementation-supported paper assets in the current RepoReviewer workflow.}
    \label{tab:paper-assets}
\end{table}

\section{Limitations}

RepoReviewer currently focuses on public repositories, depends on external language-model providers, and inherits variability from model output formatting and reasoning quality. Review quality can degrade on repositories with sparse documentation, unusual layouts, or extremely large codebases. The current report also stops short of a completed benchmark evaluation, so it should be read as an architecture and implementation paper rather than a validated empirical comparison study.

Additional limitations include provider quota and rate limits that can disrupt large-batch evaluation, severity labels that remain heuristic rather than formally calibrated, best-effort line mapping that may miss deeper cross-file issues, and the absence of private-repository support and GitHub-native review publishing in the current version.

\section{Related Work}

RepoReviewer sits at the intersection of automated software engineering, code-generation quality analysis, and agentic development systems. Benchmarks such as SWE-bench highlight the difficulty of real-world repository tasks by requiring issue resolution over complete codebases rather than isolated snippets \cite{swebench2023}. Agentic systems such as SWE-agent extend this direction by giving language models structured interfaces to repositories, files, and execution environments \cite{sweagent2024}. RepoReviewer shares the intuition that software-engineering tasks benefit from explicit structure, but it focuses on review and prioritization rather than autonomous issue resolution.

Related work on automated repair also motivates the use of staged reasoning and feedback. VRpilot studies reasoning and validation feedback in vulnerability repair and reports that decomposition and iterative refinement are important in difficult code tasks \cite{vrpilot2024}. Our work adopts a similar systems perspective, but targets repository review rather than patch synthesis.

The quality of LLM-generated code and feedback is also constrained by hallucination and reliability issues. CodeMirage documents multiple categories of hallucination in generated code, including logical and robustness failures \cite{codemirage2024}. These observations motivate our emphasis on structured outputs, deduplication, and human evaluation rather than treating model output as authoritative.

Finally, research on automated code change generation shows that practical developer tools can still be valuable without claiming fundamental model novelty. For example, PyCraft combines LLMs with transformation-by-example workflows to automate repetitive code changes in real projects \cite{pycraft2024}. More recently, ResearchPilot describes a local-first multi-agent system for literature synthesis and related-work drafting, reinforcing the value of transparent engineering architectures for domain-specific research assistance \cite{researchpilot2026}. RepoReviewer follows a comparable engineering philosophy: its contribution is the workflow design, developer-facing system integration, and reproducible evaluation scaffolding, not a new base model.

\section{Availability}

The implementation is available as an open-source repository at \url{https://github.com/peng1z/RepoReviewer}. The repository includes the Python backend, Next.js frontend, screenshots, and paper-supporting assets described in this report. The project exposes a CLI, a FastAPI backend, and a Next.js dashboard, along with templates for future empirical evaluation and annotation.

\section{Conclusion}

RepoReviewer demonstrates a practical engineering pattern for repository-level automated review. Its main contribution is architectural rather than algorithmic: decomposing review into context, analysis, prioritization, and summarization stages can improve the structure and usability of outputs while preserving a local-first developer workflow. The current report documents the system design, implementation choices, and paper-supporting infrastructure. The next step is a larger human-annotated empirical study using the included evaluation tooling.

\end{document}